# On the Necessary and Sufficient Condition of Greedy Routing Supporting Geographical Data Networks

M. Ghaffari, B. Hariri and S. Shirmohammadi

*Large scale decentralized communication systems have introduced the new trend towards online routing where routing decisions are performed based on a limited and localized knowledge of the network. Geometrical greedy routing has been among the simplest and most common online routing schemes. A perfect geometrical routing scheme is expected to deliver each packet to the point in the network that is closest to the packet destination. However greedy routing fails to guarantee such delivery as the greedy forwarding decision sometimes leads the packets to localized minimums. This article investigates the necessary and sufficient properties of the greedy supporting graphs that provide the guaranteed delivery of packets when acting as a routing substrate.*

*Introduction*: Distributed routing has become a new trend in a large variety of communication architectures ranging from ad-hoc and wireless sensor networks to massively multi-user P2P networks such as MMVEs, MMOGs, etc. In many cases, complete knowledge about the environment in which routing takes place is not available beforehand, and the path must be determined through a localized routing decision process without having the global knowledge. This family of memory-less routing schemes are generally referred to as online routing algorithms [1].

Greedy routing is known as one of the simplest and most commonly used forms of online routing. For a graph $G = (V, E)$ on the node set $V$, greedy routing tries to select the closest available node to the destination as the next hop for the forwarding packet on the condition that it finds a closer neighbour to the destination than itself.

Therefore, greedy routing decision only requires the location information of the current node and its neighbours as well as the destination point. A comprehensive survey of greedy routings in the context of Ad Hoc Networks has been provided in [2] where they have also been compared with the other types of position-based routings.

The main shortcoming of greedy routing algorithm is the probability of getting stuck at a local optimum rather than a global one. The graph $G = (V, E)$ is considered to support greedy routing *iff* greedy routing on $G$ delivers each and every packet to the closest member of $V$ to the packet's destination. It should be noted that if the destination of a packet exists in $V$, greedy routing must deliver the packet to the exact destination. According to this definition, greedy routing can also be used for multicasting to the closest neighbourhood of a desired point. This is considered as the main issue in geocasting problems as well as a common operation in many internet applications, including server selection, node clustering and peer-to-peer overlay networks [3].

Delaunay graph has been proved to provide a promising substrate for greedy routing: Dobkin et al. [4] showed that shortest path in Delaunay Triangulations is within a constant time of the shortest path in the complete graph. Bose et al. [1] and Lee et al. [3] proved that Delaunay Triangulation supports greedy routing. This article addresses the problem of greedy routing support over a graph. Here, we prove that in fact, containing Delaunay Triangulation (without degenerate edges) is a necessity and sufficient condition for supporting greedy routing. Let us first presents some notifications and definitions that will be used in the course of this work.

*Definitions*:

*1-Voronoi Cell:* Consider a set of points $V$, in the Euclidian plane. For each point $v_i \in V$, the Voronoi Cell of $v_i$, $VC(v_i)$, is defined as the set of all points in the plane that are closer to $v_i$ than any other point in *V*. It should be noted that according to this definition, Voronoi Cells of the members of *V*, partition the Euclidian plane.

*2-Delaunay Triangulation:* Presuming a set of points V, Delaunay Triangulation of V, is a graph G= (V, E) where e= ($v_i$, $v_j$) ∈ E, *iff* VC($v_i$) and VC($v_j$) has a side (or at least a point) in common. If VC(vi) and VC(vj) only share a single point, the edge e= ($v_i$, $v_j$) is referred to as a degenerate edge. Delaunay Triangulation is a dual of Voronoi diagram. An example of Voronoi diagram and Delaunay triangulation is represented in Fig.1.

*3-Vertex Region:* Consider a graph G= (V, E) on the set of points V, where for each $v_i$ ∈ V, N($v_i$) stands for the neighbour set of $v_i$. For each $v_i$ ∈ V, we define the Vertex Region of $v_i$ in graph G, $VR_G(v_i)$, as the set of all points in the plane that are closer to $v_i$ than to any other point in N($v_i$). Fig. 2 shows the Vertex Region for a node of a sample graph in long with the Voronoi diagram of the same set of nodes.

*4-Delaunay Graph: Having a set of points V in the plane, we construct a Graph G= (V, E) on this set in a way that E is the set of all non-degenerate edges of DT(V). We call this graph, Delaunay graph of the set of points.*

***Remark 1:*** For each $v_i$ ∈ V, we have N($v_i$) ⊂ V and thus, VC($v_i$) ⊂ $VR_G(v_i)$.

*Greedy routing supporting graphs:*

***Theorem 1:*** Graph G= (V, E) on the nodes V in the Euclidian plane supports greedy routing *iff* for every node $v_i$ ∈ V, $VR_G(v_i)$ = VC($v_i$).

***Proof.*** For the sufficiency condition, suppose that the graph G=(V, E) satisfies the condition $VR_G(v_i)$ = VC($v_i$) for each node $v_i$ ∈ V. We can then prove that G supports greedy routing by contradiction: Assuming that G does not support greedy routing, there exists a point $v_i$ ∈ V where a packet might get stuck and cannot be forwarded any further while $v_i$ is not the nearest point in V to the packet destination. Since $v_i$ is not the nearest point in V to the packet destination, packet destination is located outside of the $VC(v_i)$ and as $VR_G(v_i) = VC(v_i)$, the destination point is also located out of $VR_G(v_i)$. Therefore, there should be a point $v_j \in N(v_i)$ that is closer to the packet destination than $v_i$ to which the packet will be forwarded. This is in contradiction with the assumption of the packet getting stuck in $v_i$.

For the necessity condition, it is assumed that G supports greedy routing. We will then prove that $VR_G(v_i) = VC(v_i)$ for every $v_i \in V$. It should be noted that according to *Remark 1*, for each $v_i \in V$, $VC(v_i) \subset VR_G(v_i)$. Therefore, it suffices to show that $VR_G(v_i) \subset VC(v_i)$, as well. This can be proven by contradiction. Suppose that there exists a point $x \in VR_G(v_i)$ that does not belong to $VC(v_i)$. Since $x \notin VC(v_i)$ and Voronoi Cells partition the plane, x is in the Voronoi Cell of another node $v_j$. Therefore, $v_i$ is not the closest member of V to x. On the other side $x \in VR_G(v_i)$, and hence, $v_i$ has no closer neighbour to x than itself. Thus, if a packet destined to x reaches $v_i$ or starts in $v_i$, it cannot be delivered to $v_j$. Therefore G does not support greedy routing and this is in contradiction with the initial assumption. □

***Theorem 2:*** Graph G= (V, E) supports greedy routing ***iff*** E contains all of non-degenerate edges of DT(V) as an its sub-set.

***Proof.*** It is obvious that if E contains all of non-degenerate edges of DT(V) as its sub-set, we will have $VR_G(v_i) = VC(v_i)$ for every $v_i \in V$ and thus, using Theorem 1, it is concluded that G supports greedy routing.

We prove the necessity condition by contradiction. Suppose that G supports Greedy routing and there exists a non-degenerate edge, $e= (v_i, v_j)$, in DT(V) that does not belong to *E*. As *e* is a non-degenerate edge of DT(V), $VC(v_i)$ and $VC(v_j)$ should have at least two different points: $p_1, p_2$ in common. Therefore $p_1, p_2 \in VC(v_i) \cap VC(v_j)$ and $p_1 \neq p_2$. Assume that p is the middle point of the line connecting $p_1$ to $p_2$. Obviously, *p*

$\in VC(v_i) \cap VC(v_j)$ and thus, *p* is a boundary point of $VC(v_i)$. As $p \in VC(v_i)$, it can be concluded that $p \in VR_G(v_i)$ according to *Remark1*. Thus, p is either an interior or a boundary point of $VR_G(v_i)$. We show that p is an interior point of $VR_G(v_i)$ and thus, $VR_G(v_i) \neq VC(v_i)$ which is in contradiction with the assumption that G supports greedy routing, using *Theorem 1*. If p is assumed to be a boundary point of $VR_G(v_i)$, $v_i$ has a neighbour $v_k$ such that $d(p,v_i) = d(p,v_k)$, where *d* exhibits the Euclidian distance. Therefore, $v_k$ is on the circle centred at p and passing $v_i$ (Fig. 3). As $v_k \neq v_j$, orthogonal bisector of line segment connecting $v_i$ to $v_k$ crosses the orthogonal bisector of line segment from $v_i$ to $v_j$. This has been demonstrated in Fig. 3. Therefore, points $p_1$ and $p_2$ belong to different half-planes constructed by orthogonal bisector of line segment from $v_i$ to $v_k$. Hence, one of these points is strictly closer to $v_k$ than $v_i$. However, this is in contradiction with the assumption that $p_1$ and $p_2 \in VC(v_i)$. □

*Theorem 2* completely specifies the characteristics of greedy routing supporting geometric graphs. According to *Theorem 2*, any graph G on the point set V supports greedy routing *iff,* G contains Delaunay Graph of V as an its sub-graph. Therefore, Delaunay Graph of V is the sparsest greedy supporting graph on V. It should be noted that Delaunay Graph is usually the same as Delaunay Triangulation while in exceptional cases Delaunay Triangulation contains a number of de-generate edges. It is easy to see that this exceptional condition occurs if and only if there are more than 3 points of V on the same circle. However, in this condition the unique definition of Delaunay triangulation fails and there may be some Delaunay triangulations, for the same set of points. In the other words, we can say non-degenerate edges of a Delaunay triangulation are the sure edges of any Delaunay triangulation.

*Conclusion:*

In this article, we have investigated the properties of greedy supporting graphs and proven that containing Delaunay graph as sub-graph (containing all non-degenerate edges of DT as sub-set of edges) is a necessity and sufficient condition for supporting greedy routing.

**Authors' affiliations:**

M. Ghaffari and B. Hariri (Department of Electrical Engineering, Sharif University of Technology, Tehran, Iran). {Ghaffari | Hariri}@ee.sharif.edu}
B. Hariri and S. Shirmohammadi (Distributed Collaborative Virtual Environment Research Laboratory, University of Ottawa, Ottawa, Canada). {BHariri | Shervin}@discover.uottawa.ca

**Figure captions:**

Fig. 1 Voronoi Diagram and Delaunay Triangulation
─ . ─ . ─ Voronoi Diagram
──────── Delaunay Triangulation

Fig. 2 Vertex Region Definition
─ . ─ . ─ Voronoi Diagram
────── Example Graph
─ ─ ─ ─ Vertex Region of one node

Fig. 3 Proof of *Theorem 2*
– . – . – Orthogonal bisector of line segment $v_i$ to $v_j$
– – – – Vertex Regions of $v_i$ and $v_j$

Figure 1

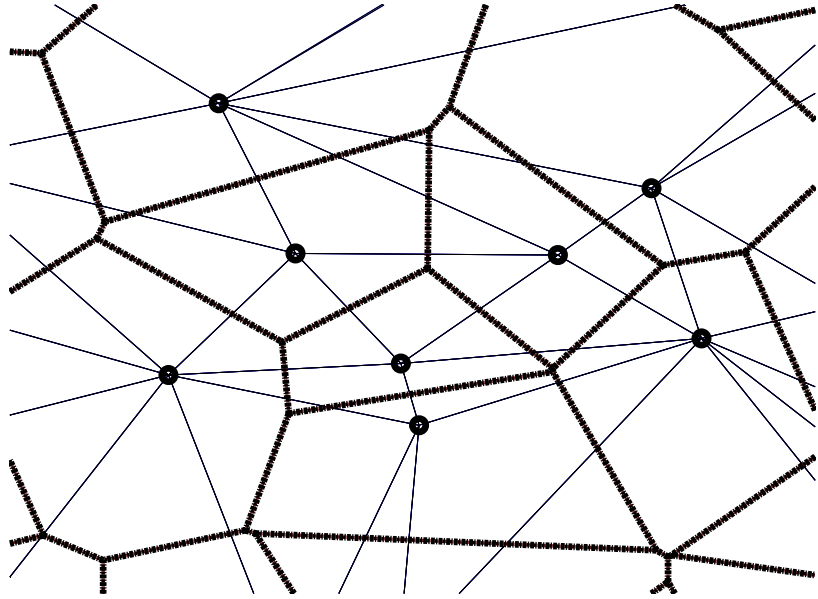

Figure 2

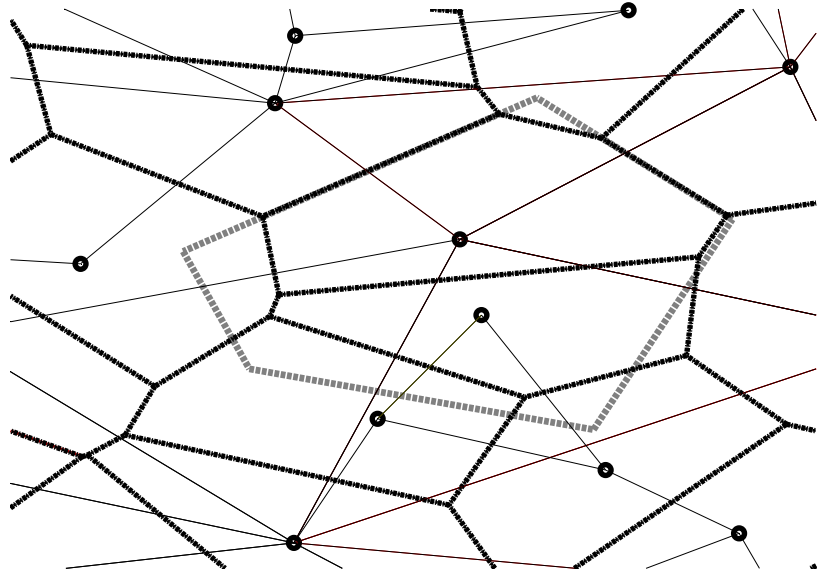

Figure 3

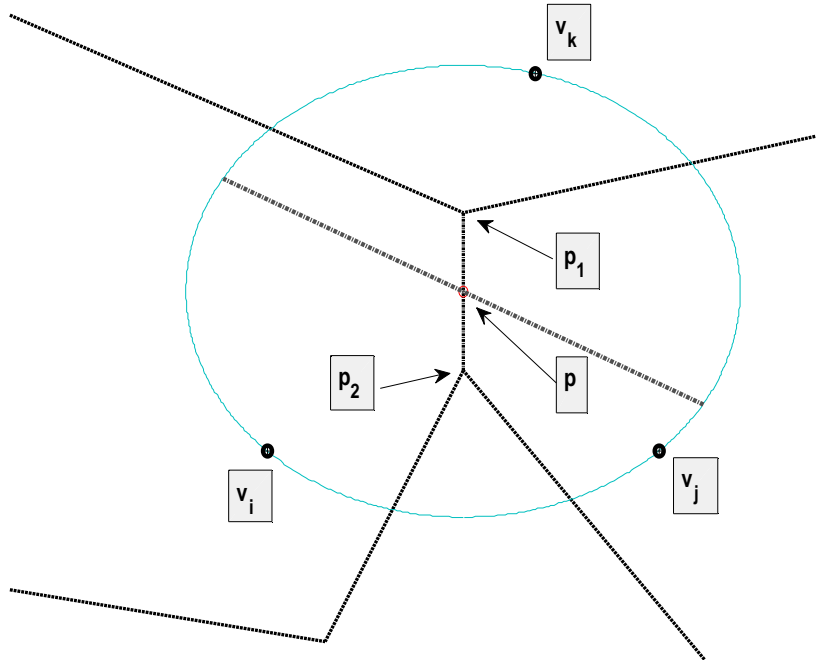